\documentclass[a4paper]{jpconf}

\usepackage{bm}

\begin{document}

\title{{\bf Weyl and Majorana for Neutral Particles}}

%\footnote{Talk at the Conference General 
%Relativity, Quantumm Mechanics and Everything 
%in Between Celebrating 02 Spring of Prof. Horwitz. Tel Aviv, Israel.}}

\address{{\bf Valeriy V. Dvoeglazov}\\ UAF, Universidad Aut\'onoma de Zacatecas, M\'exico\\E-mail: valeri@fisica.uaz.edu.mx}

\date{\empty}

%\maketitle

\begin{abstract}
We compare various formalisms for neutral particles. It is found that they contain unexplained contradictions.
Next, we investigate the spin-1/2 and spin-1 cases in different bases. Next, we look for relations 
with the Majorana-like field operator. We show explicitly incompatibility of the Majorana anzatzen with the Dirac-like field operators in both the original Majorana theory and its generalizations. Several explicit examples are presented for higher spins too. It seems that the calculations in the helicity basis only give mathematically and physically reasonable results.
\end{abstract}

\large{
\section{Weyl Formalism.}

The Weyl formalism is just a massless limit of the Dirac equation:
\begin{equation}
[i\gamma^\mu \partial_\mu ]\Psi (x) =0\,.
\end{equation}
Of course, it can be re-written in the 2-component forms:
\begin{eqnarray}\
[ p_0 + {\bf \sigma}\cdot  {\bf p} ] \chi (x) = 0\,,\quad
[ p_0 - {\bf \sigma}\cdot  {\bf p} ] \phi (x) = 0\,.
\end{eqnarray}
However, if we apply the Noether theorem to the Lagrangian
\begin{equation}
{\cal L}  = {i\over 2} [\bar \Psi \gamma_\mu\partial^\mu
 \Psi - \partial^\mu\bar \Psi \gamma_\mu  \Psi ]
\end{equation}
we obtain the current operator
\begin{equation}
J_\mu = \bar \Psi \gamma_\mu \Psi \,,
\end{equation}
as in the massive case.
So, it is doubtful that we can use the massless limit of the Dirac equation for neutral particles.

%\section{Introduction.}

In Refs.~\cite{DV2}-\cite{DVCONF2} we considered the procedure of construction of the field operators {\it ab initio} 
(including for neutral particles). The Bogoliubov-Shirkov method has been used.

In the present article we investigate the spin-1/2 and spin-1 cases in different bases. The Majorana theory of the neutral particles is well known~\cite{Majorana}. We look for relations of the Dirac-like field operator 
to the Majorana-like field operator. It seems that the calculations in the helicity basis give mathematically and physically reasonable results.

%\newpage

\section{The Spin-1/2.}

Usually, everybody uses the following definition of the field operator~\cite{Itzyk} in the pseudo-Euclidean metrics:
\begin{equation}
\Psi (x) = \frac{1}{(2\pi)^3}\sum_h \int \frac{d^3 {\bf p}}{2E_p} [ u_h ({\bf p}) a_h ({\bf p}) e^{-ip_\mu\cdot x^\mu}
+ v_h ({\bf p}) b_h^\dagger ({\bf p}) e^{+ip_\mu\cdot x^\mu}]\,,
\end{equation} 
as given {\it ab initio}.
The momentum-space 4-spinors ( $u-$ and $v-$) 
satisfy the equations: $(\hat p - m) u_h (p) =0$ and $( \hat p + m) v_h (p) =0$, respectively; the $h$ is 
the polarization index. It is easy to prove from the characteristic equations
$Det (\hat p \mp m) =(p_0^2 -{\bf p}^2 -m^2)^2= 0$ that the solutions should satisfy the energy-momentum relations $p_0= \pm E_p =\pm \sqrt{{\bf p}^2 +m^2}$ for both $u-$ and $v-$ solutions.

The general scheme of construction of the field operator has been given in~\cite{Bogoliubov}. In the case of
the $(1/2,0)\oplus (0,1/2)$ representation we have:
\begin{equation}
\Psi (x) = \frac{1}{(2\pi)^{3}}\int dp\,  e^{ip\cdot x} \tilde \Psi (p)\,.
\end{equation}
We know the condition of the mass shell: $(p^2 -m^2) \tilde \Psi (p) =0$.
Thus, $\tilde \Psi (p) = \delta (p^2 - m^2) \Psi (p)$. After simple transformations
we  obtain
\begin{eqnarray}\
\Psi (x)&=& {1\over (2\pi)^3} \sum_h^{} \int {d^3 {\bf p} \over 2E_p} \theta(p_0)  
\left [ u_h (p) a_h (p)\vert_{p_0=E_p} e^{-i(E_p t-{\bf p}\cdot {\bf x})}+ \right. \nonumber\\
&+& \left. u_h (-p) a_h (-p)\vert_{p_0=E_p} e^{+i (E_p t- {\bf p}\cdot {\bf x})} 
\right ]\nonumber
\end{eqnarray}
During the calculations we had to represent $1=\theta (p_0) +\theta (-p_0)$ above
in order to get positive- and negative-frequency parts.
We did not yet assumed, which equation does this
field operator  (namely, the $u-$ spinor) satisfy, with negative- or positive- mass and/or $p^0= \pm E_p$.
We should transform $u_h (-p)$ to the $v_h (p)$ 4-spinor. The procedure is the following one~\cite{DV1,DV2}.
%\subsection{The Standard Basis.}
 
In the Dirac case we should assume the following relation in the field operator:
\begin{equation}
\sum_{h=\pm 1/2}^{} v_h (p) b_h^\dagger (p) = \sum_{h=\pm 1/2}^{} u_h (-p) a_h (-p)\,,\label{dcop}
\end{equation}
which is compatible with the ``hole" theory and the Feynman-Stueckelberg interpretation.
We need $\Lambda_{\mu\lambda} (p) = \bar v_\mu (p) u_\lambda (-p)$.
By direct calculations,  we find
\begin{equation}
-mb_\mu^\dagger (p) = \sum_{\lambda}^{} \Lambda_{\mu\lambda} (p) a_\lambda (-p)\,.
\end{equation}
Hence, $\Lambda_{\mu\lambda} = - im ({\bm \sigma}\cdot {\bf n})_{\mu\lambda}$, ${\bf n} = {\bf p}/\vert{\bf p}\vert$, 
and 
\begin{equation}
b_\mu^\dagger (p) = + i\sum_\lambda ({\bm\sigma}\cdot {\bf n})_{\mu\lambda} a_\lambda (-p)\,.\label{co}
\end{equation}
Multiplying (\ref{dcop}) by $\bar u_\mu (-p)$ we obtain
\begin{equation}
a_\mu (-p) = -i \sum_{\lambda} ({\bm \sigma} \cdot {\bf n})_{\mu\lambda} b_\lambda^\dagger (p)\,.\label{ao}
\end{equation}
The equations are self-consistent.

%\subsection{The Helicity Basis.}

The details of the helicity basis are given in Refs.~\cite{GR,DVIJTP}.
However, in this helicity case we have:
\begin{equation}
\Lambda_{hh^\prime} (p) = \bar v_h (p) u_{h^\prime} (-p)=
i\sigma^y_{hh^\prime}\,.
\end{equation}
So, someone  may argue that we should introduce the creation operators by hand in every basis.

%\subsection{Application of the Majorana anzatzen.}

It is well known that  ``{\it particle=antiparticle}" in the Majorana theory~\cite{Majorana}. So, in the language of the quantum field theory we should have: 
\begin{equation}
b_\mu (E_p, {\bf p}) = e^{i\varphi} a_\mu (E_p, {\bf p})\,.\label{ma}
\end{equation} 
Usually, different authors use $\varphi = 0, \pm \pi/2$ depending on the metrics and on the forms of the 4-spinors and commutation relations.  So, on using (\ref{co}) and the above-mentioned postulate we come to:
\begin{equation}
a_\mu^\dagger (p) = +i e^{i\varphi} ({\bm \sigma}\cdot {\bf n})_{\mu\lambda} a_\lambda (-p)\,.\label{ma1}
\end{equation}
On the other hand, on using (\ref{ao}) we make the substitutions $E_p \rightarrow -E_p$, ${\bf p} 
\rightarrow -{\bf p}$ to obtain 
\begin{equation}
a_\mu (p) = +i ({\bm \sigma}\cdot {\bf n})_{\mu\lambda} b_\lambda^\dagger (-p)\,.\label{ma2}
\end{equation}
The totally reflected (\ref{ma}) is $b_\mu (-E_p, -{\bf p}) = e^{i\varphi} a_\mu (-E_p, -{\bf p})$.
Thus,
\begin{equation}
b_\mu^\dagger (- p) = e^{-i\varphi} a_\mu^\dagger (- p)\,.
\end{equation}
Combining with (\ref{ma2}),  we come to
\begin{equation}
a_\mu (p) = +i e^{-i\varphi}({\bm \sigma}\cdot {\bf n})_{\mu\lambda} a_\lambda^\dagger (-p)\,,
\end{equation}
and 
\begin{equation}
a_\mu^\dagger (p) = -i e^{i\varphi}({\bm \sigma}^\ast \cdot {\bf n})_{\mu\lambda} a_\lambda (-p)\,.
\end{equation}
This contradicts with the above equation  unless we have the preferred axis in every inertial system.

Next, we can use another Majorana anzatz $\Psi = \pm e^{i\alpha} \Psi^{c}$ with usual definitions
\begin{eqnarray}
{\cal C} =e^{i\vartheta_c}\pmatrix{0&i\Theta\cr -i\Theta & 0\cr} {\cal K}\,,\quad
\Theta = \pmatrix{0&-1\cr 1&0\cr}\,.\label{cco}
\end{eqnarray}
Thus, on using $Cu_\uparrow^\ast ({\bf p}) =iv_\downarrow ({\bf p})$, $Cu_\downarrow^\ast ({\bf p}) =-iv_\uparrow ({\bf p})$ we come to other relations
between creation/annihilation operators 
\begin{eqnarray}
a_\uparrow^\dagger  ({\bf p}) &=& \mp i e^{-i\alpha} b_\downarrow^\dagger ({\bf p})\,,\\
a_\downarrow^\dagger  ({\bf p}) &=& \pm i e^{-i\alpha} b_\uparrow^\dagger ({\bf p})\,,
\end{eqnarray}
which may be used instead of (\ref{ma}).
Due to the possible signs $\pm$ the number of the corresponding states is the same as in the Dirac case that permits us to have the complete system 
of the Fock states over the $(1/2,0)\oplus (0,1/2)$ representation space in the mathematical sense.\footnote{Please note that the phase factors may have physical significance in quantum field theories as opposed to the textbook nonrelativistic quantum mechanics, as was discussed recently by several authors.}
However, in this case we deal with the self/anti-self charge conjugate quantum field operator instead of the self/anti-self charge conjugate quantum states. Please remember that it is the latter that answers for the neutral particles.

We conclude that something is missed in the foundations of both the original Majorana theory and its generalizations
in the $(1/2,0)\oplus (0,1/2)$ representation.

%\subsection{Self/Anti-self Charge Conjugate States.}

We  define the {\it self/anti-self} charge-conjugate 4-spinors 
in the momentum space~\cite{DVA1995}:
\begin{eqnarray}
{\cal C}\lambda^{S,A} ({\bf p}) &=& \pm \lambda^{S,A} ({\bf p})\,,\\
{\cal C}\rho^{S,A} ({\bf p}) &=& \pm \rho^{S,A} ({\bf p})\,.
\end{eqnarray}
Such definitions of 4-spinors differ, of course, from the original Majorana definition in x-representation:
\begin{equation}
\nu (x) = \frac{1}{\sqrt{2}} (\Psi_D (x) + \Psi_D^c (x))\,,
\end{equation}
$C \nu (x) = \nu (x)$ that represents the positive real $C-$ parity field operator. However, the momentum-space Majorana-like spinors 
open various possibilities for description of neutral  particles 
(with experimental consequences, see~\cite{Kirchbach}). For instance, ``for imaginary $C$ parities, the neutrino mass 
can drop out from the single $\beta $ decay trace and 
reappear in $0\nu \beta\beta $, a curious and in principle  
experimentally testable signature for a  non-trivial impact of 
Majorana framework in experiments with polarized sources."

Thus, in the accustomed basis the explicit forms of the 4-spinors of the second kind  $\lambda^{S,A}_{\uparrow\downarrow}
({\bf p})$ and $\rho^{S,A}_{\uparrow\downarrow} ({\bf p})$
are:
\begin{eqnarray}
\lambda^S_\uparrow ({\bf p}) &=& \frac{1}{2\sqrt{E_p+m}}
\pmatrix{ip_l\cr i (p^- +m)\cr p^- +m\cr -p_r},
\lambda^S_\downarrow ({\bf p})= \frac{1}{2\sqrt{E_p+m}}
\pmatrix{-i (p^+ +m)\cr -ip_r\cr -p_l\cr (p^+ +m)}\nonumber\\
\\
\lambda^A_\uparrow ({\bf p}) &=& \frac{1}{2\sqrt{E_p+m}}
\pmatrix{-ip_l\cr -i(p^- +m)\cr (p^- +m)\cr -p_r},
\lambda^A_\downarrow ({\bf p}) = \frac{1}{2\sqrt{E_p+m}}
\pmatrix{i(p^+ +m)\cr ip_r\cr -p_l\cr (p^+ +m)}\nonumber
\end{eqnarray}
In this basis one has
\begin{eqnarray}
\rho^S_\uparrow ({\bf p}) \,&=&\, - i \lambda^A_\downarrow ({\bf p})\,,\,
\rho^S_\downarrow ({\bf p}) \,=\, + i \lambda^A_\uparrow ({\bf p})\,,\,\\
\rho^A_\uparrow ({\bf p}) \,&=&\, + i \lambda^S_\downarrow ({\bf p})\,,\,
\rho^A_\downarrow ({\bf p}) \,=\, - i \lambda^S_\uparrow ({\bf p})\,.
\end{eqnarray}

The $\lambda -$ and $\rho -$ spinors are connected with the $u-$ and $v-$ spinors by the  following formula:
\begin{eqnarray}
\pmatrix{\lambda^S_\uparrow ({\bf p}) \cr \lambda^S_\downarrow ({\bf p}) \cr
\lambda^A_\uparrow ({\bf p}) \cr \lambda^A_\downarrow ({\bf p})\cr} = {1\over
2} \pmatrix{1 & i & -1 & i\cr -i & 1 & -i & -1\cr 1 & -i & -1 & -i\cr i&
1& i& -1\cr} \pmatrix{u_{+1/2} ({\bf p}) \cr u_{-1/2} ({\bf p}) \cr
v_{+1/2} ({\bf p}) \cr v_{-1/2} ({\bf p})\cr},\label{connect}
\end{eqnarray}
provided that the 4-spinors have the same physical dimension.

We construct the field operators on using the procedure above with $\lambda^S_\eta (p)$.
Thus, the difference is that 1) instead of
$u_h (\pm p)$ we have $\lambda^S_\eta (\pm p)$; 2) possible change of the annihilation operators,
$a_h \rightarrow c_\eta$. Apart, one can make corresponding changes due to 
normalization factors. Thus, we should have 
\begin{equation}
\sum_{\eta=\pm 1/2}^{} \lambda^A_\eta (p) d_\eta^\dagger (p) = \sum_{\eta=\pm 1/2}^{} \lambda^S_\eta (-p) c_\eta (-p)\,.\label{dcop1}
\end{equation}
We find surprisingly:
\begin{equation}
d^\dagger_\eta (p) = -\frac{ip_y}{p}\sigma^y_{\eta\tau}  c_\tau (-p)\,,\quad
c_\eta (-p) = -\frac{ip_y}{p}\sigma^y_{\eta\tau}  d^\dagger_\tau (p)\,.
\end{equation}
The bi-orthogonal anticommutation relations are given in Ref.~\cite{DVA1995}. See other details in Ref.~\cite{DVNCA,DVMPLA}.
Concerning with the $P$,$C$ and $T$ properties of the corresponding states see Ref.~\cite{DVMPLA} in this model.

The above-mentioned contradictions may be related to the possibility of the conjugation which is different from that of Dirac.
Both in the Dirac-like case and the Majorana-like case ($c_\eta (p)= e^{-i\varphi} d_\eta (p)$) we have difficulties in the construction of field operators.

\section{The Spin-1.}

%\subsection{The Standard Basis.}

We use the results of Refs.~\cite{Novozh,Wein,Dv-book} in this Section.
The polarization vectors of the standard basis are defined~\cite{Var}:
\begin{eqnarray}
&&\epsilon^\mu
({\bf 0}, +1)= -{1\over \sqrt{2}}\pmatrix{0\cr 1\cr i\cr 0 \cr}\,,\quad  
\epsilon^\mu ({\bf 0}, -1)= +{1\over
\sqrt{2}}\pmatrix{0\cr 1\cr -i\cr 0\cr}\,,\\
&&\epsilon^\mu ({\bf
0}, 0) = \pmatrix{0\cr 0\cr 0\cr 1\cr}\,,\quad
\epsilon^\mu ({\bf
0}, 0_t) = \pmatrix{1\cr 0\cr 0\cr 0\cr}\,.
\end{eqnarray}
The Lorentz transformations are ($\widehat p_i = p_i/\vert {\bf p}\vert$):
\begin{eqnarray} 
&& \epsilon^\mu ({\bf p}, \sigma) =
L^{\mu}_{\,\,\nu} ({\bf p}) \epsilon^\nu ({\bf 0},\sigma)\,,\\ 
&& L^{0}_{\,\,0} ({\bf p}) = \gamma\, ,\, L^{i}_{\,\,0} ({\bf p}) = 
L^{0}_{\,\,i} ({\bf p}) = \widehat p_i \sqrt{\gamma^2 -1}\, ,\,
L^{i}_{\,\,k} ({\bf p}) = \delta_{ik} + (\gamma -1) \widehat p_i \widehat
p_k \,.\nonumber\\ 
\end{eqnarray}
Hence, for the particles of the mass $m$ we have:
\begin{eqnarray} 
u^\mu
({\bf p}, +1)&=& -{N\over \sqrt{2}m}\pmatrix{-p^r\cr m+ {p^1 p^r \over
E_p+m}\cr im +{p^2 p^r \over E_p+m}\cr {p^3 p^r \over
E_p+m}\cr}\,,\quad 
u^\mu ({\bf p}, -1)= {N\over
\sqrt{2}m}\pmatrix{-p^l\cr m+ {p^1 p^l \over E_p+m}\cr -im +{p^2 p^l \over
E_p+m}\cr {p^3 p^l \over E_p+m}\cr}\,,\nonumber\\
&&\label{vp12}\\ u^\mu ({\bf
p}, 0) &=& {N\over m}\pmatrix{-p^3\cr {p^1 p^3 \over E_p+m}\cr {p^2 p^3
\over E_p+m}\cr m + {(p^3)^2 \over E_p+m}\cr}\,,\quad
u^\mu ({\bf p}, 0_t) = {N \over m} \pmatrix{E_p\cr -p^1
\cr -p^2\cr -p^3\cr}\,.
\end{eqnarray}
$N$ is the normalization constant for $u^\mu ({\bf p},\sigma)$. 
They are the eigenvectors of the parity operator ($\gamma_{00}= \mbox{diag} (1\,\, -1\,\, -1\,\, -1)$):
\begin{equation}
\hat P u_\mu (-{\bf p}, \sigma) = - u_\mu ({\bf p}, \sigma)\,,\quad
\hat P u_\mu (-{\bf p}, 0_t) = +u_\mu ({\bf p}, 0_t)\,.
\end{equation}
It is assumed that they form the complete orthonormalized system of the $(1/2,1/2)$
represntation, $\epsilon_\mu^\ast ({\bf p}, 0_t) \epsilon^\mu ({\bf p}, 0_t)=1$,
$\epsilon_\mu^\ast ({\bf p}, \sigma^\prime ) \epsilon^\mu ({\bf p}, \sigma)=-\delta_{\sigma^\prime \sigma}$.

%\subsection{The Helicity Basis.}
%~\cite{GR}
The helicity operator should act as:
\begin{equation}
%{({\bf S}\cdot {\bf p})\over p} = {1\over p} \pmatrix{0&0&0&0\cr
%0&0&-ip^3&ip^2\cr
%0&ip^3&0&-ip^1\cr
%0&-ip^2&ip^1&0\cr}\,,\,\,
{({\bf S}\cdot {\bf p})\over p} \epsilon^\mu_{\pm 1} = \pm \epsilon^\mu_{\pm 1}\,,\,\,{({\bf S}\cdot {\bf p})\over p} \epsilon^\mu_{0,0_t} = 0\,.
\end{equation}
The eigenvectors are in the helicity basis:
\begin{eqnarray}
&&\epsilon^\mu_{+1}= {1\over \sqrt{2}} {e^{i\alpha}\over p} \pmatrix{0\cr
{-p^1 p^3 +ip^2 p\over \sqrt{(p^1)^2 +(p^2)^2}}\cr {-p^2 p^3 -ip^1
p\over \sqrt{(p^1)^2 +(p^2)^2}}\cr \sqrt{(p^1)^2 +(p^2)^2}\cr}\,,\,
\epsilon^{\mu }_{-1}={1\over \sqrt{2}}{e^{i\beta}\over p}
\pmatrix{0\cr {p^1 p^3 +ip^2 p\over \sqrt{(p^1)^2 +(p^2)^2}}\cr {p^2 p^3
-ip^1 p\over \sqrt{(p^1)^2 +(p^2)^2}}\cr -\sqrt{(p^1)^2 +(p^2)^2}\cr} \nonumber\\
&&\\
&&\epsilon^{\mu }_0={\frac{1}{m}}\pmatrix{p\cr {E \over p}
p^1 \cr {E \over p} p^2 \cr{E \over p} p^3\cr}\,,\quad
\epsilon^{\mu }_{0_{t}}={\frac{1}{m}}\pmatrix{E_p\cr p^1\cr
p^2\cr p^3\cr}\,.
\end{eqnarray}
The normalization is the same as in the standard basis.
The eigenvectors $\epsilon^\mu_{\pm  1}$ are not the eigenvectors
of the parity operator ($\gamma_{00} R$) of this representation. However, the $\epsilon^\mu_{1,0}$,
$\epsilon^\mu_{0,0_t}$
are. 
Various-type field operators are possible in this representation. Let us remind the procedure to get them.
\begin{eqnarray}
&&A_\mu (x) = {1\over (2\pi)^3} \int d^4 p \,\delta (p^2 -m^2) e^{-ip\cdot x}
A_\mu (p) =\nonumber\\
&=&{1\over (2\pi)^3} \sum_{\lambda}^{}\int {d^3 {\bf p} \over 2E_p}   
[\epsilon_\mu (p,\lambda) a_\lambda (p) e^{-ip\cdot x}  + \epsilon_\mu (-p,\lambda) 
a_\lambda (-p) e^{+ip\cdot x} ]\,.
\end{eqnarray}
We should transform the second part to $\epsilon_\mu^\ast (p,\lambda) b_\lambda^\dagger (p)$ as usual. In such a way we obtain the states which are considered to be the charge-conjugate 
states. In this Lorentz group representation the charge conjugation operator is just the complex conjugation operator for 4-vectors. We postulate
\begin{equation}
\sum_{\lambda}^{} \epsilon_\mu (-p,\lambda) a_\lambda (-p) = 
\sum_{\lambda}^{} \epsilon_\mu^\ast (p,\lambda) b_\lambda^\dagger (p)\,.
\label{expan}
\end{equation}
Then we multiply both parts by $\epsilon^\mu (p,\sigma)$, and use the normalization conditions for polarization vectors.

In the $({1\over 2}, {1\over 2})$ representation we can also expand
(apart of the equation (\ref{expan})) in a different way. For example,
\begin{equation}
\sum_{\lambda}^{} \epsilon_\mu (-p, \lambda) c_\lambda (-p) =
\sum_{\lambda}^{} \epsilon_\mu (p, \lambda) d^\dagger_\lambda (p)\,.
\end{equation}
From the first definition we obtain:
\begin{eqnarray}
&&\pmatrix{b^\dagger_{0_t} (p)\cr
-b^\dagger_{+1} (p)\cr
-b^\dagger_{0} (p)\cr
-b^\dagger_{-1} (p)\cr} =
 \sum_{\mu\lambda}^{} \epsilon^\mu (p,\sigma) 
\epsilon_\mu (-p,\lambda) a_\lambda (-p)= \sum_{\lambda} \Lambda_{\sigma\lambda}^{(1a)} a_\lambda (-p)=\nonumber\\ 
&=&\pmatrix{-1&0&0&0\cr
0&\frac{p_r^2}{p^2}& - \frac{\sqrt{2} p_z p_r}{p^2}& \frac{p_z^2}{p^2}\cr
0&- \frac{\sqrt{2} p_z p_r}{p^2}&-1 +\frac{2p_z^2}{p^2}& +\frac{\sqrt{2}p_z p_l}{p^2}\cr
0&\frac{p_z^2}{p^2}&+ \frac{\sqrt{2} p_z p_l}{p^2}& \frac{p_l^2}{p^2}\cr}\pmatrix{a_{00} (-p)\cr a_{11} (-p)\cr
a_{10} (-p)\cr a_{1-1} (-p)\cr}\,.
\end{eqnarray}
%or
%\begin{eqnarray}
%b_\sigma^\dagger (p) = {E_p^2 \over m^2} \pmatrix{1+{{\bf p}^2\over E_p^2}&\sqrt{2}
%{p_r \over E_p}&-\sqrt{2} {p_l \over E_p}& -{2p_3 \over E_p}\cr
%-\sqrt{2} {p_r \over E_p}&-{p_r^2 \over {\bf p}^2}& -{m^2p_3^2\over E_p^2 {\bf p}^2}
%+{p_r p_l \over E_p^2} & {\sqrt{2} p_3 p_r \over {\bf p}^2}\cr
%\sqrt{2} {p_l \over E_p}&-{m^2 p_3^2 \over E_p^2 {\bf p}^2} + {p_r p_l \over E_p^2}& -{p_l^2\over {\bf p}^2} & -{\sqrt{2} p_3 p_l \over {\bf p}^2}\cr
%{2p_3 \over E_p}&{\sqrt{2}p_3 p_r \over {\bf p}^2}& -{\sqrt{2} p_3 p_l\over {\bf p}^2} & {m^2 \over E_p^2} -{2 p_3 \over {\bf p}^2}\cr}
%\pmatrix{a_{00} (-p)\cr a_{11} (-p)\cr
%a_{1-1} (-p)\cr a_{10} (-p)\cr}.\nonumber
%\end{eqnarray}
%\begin{eqnarray}
%b^\dagger_\sigma (p) =  \pmatrix{-1&0&0&0\cr
%0&{p_3^2 \over {\bf p}^2}& {p_l^2\over {\bf p}^2} & {\sqrt{2} p_3 p_l \over {\bf p}^2}\cr
%0&{p_r^2 \over {\bf p}^2}& {p_3^2\over {\bf p}^2} & -{\sqrt{2} p_3 p_r \over {\bf p}^2}\cr
%0&{\sqrt{2}p_3 p_r \over {\bf p}^2}& -{\sqrt{2} p_3 p_l\over {\bf p}^2} & 1-{2 p_3^2 \over {\bf p}^2}\cr}\pmatrix{a_{00} (-p)\cr a_{11} (-p)\cr
%a_{1-1} (-p)\cr a_{10} (-p)\cr}\,.
%\end{eqnarray}

Possibly, we should think about modifications of the Fock space in this case. Alternatively, one can think to introduce several field operators for the $({1\over 2}, {1\over 2})$ representation. The Majorana-like anzatz is compatible for the $0_t$ time-like polarization state only in this basis of this representation.
However, the corresponding matrices $\Lambda^2$ in the helicity basis are different. Here they are:
\begin{eqnarray}
&&\pmatrix{b^\dagger_{0_t} (p)\cr
-b^\dagger_{+1} (p)\cr
-b^\dagger_{0} (p)\cr
-b^\dagger_{-1} (p)\cr} =
 \sum_{\mu\lambda}^{} \epsilon^\mu (p,\sigma) 
\epsilon_\mu (-p,\lambda) a_\lambda (-p)= \sum_{\lambda} \Lambda_{\sigma\lambda}^{(2a)} a_\lambda (-p)=\nonumber\\ 
&=&-\pmatrix{1&0&0&0\cr
0&e^{2i\alpha}&0&0\cr
0&0&1&0\cr
0&0&0&e^{2i\beta}}\pmatrix{a_{00} (-p)\cr a_{11} (-p)\cr
a_{10} (-p)\cr a_{1-1} (-p)\cr}\,,
\end{eqnarray}
and
%\end{document}
\begin{eqnarray}
&&\pmatrix{d^\dagger_{0_t} (p)\cr
-d^\dagger_{+1} (p)\cr
-d^\dagger_{0} (p)\cr
-d^\dagger_{-1} (p)\cr} =
 \sum_{\mu\lambda}^{} \epsilon^{\mu ^\ast} (p,\sigma) 
\epsilon_\mu (-p,\lambda) c_\lambda (-p)= \sum_{\lambda} \Lambda_{\sigma\lambda}^{(2b)} c_\lambda (-p)=\nonumber\\ 
&=&-\pmatrix{1&0&0&0\cr
0&0&0&-e^{-i(\alpha-\beta)}\cr
0&0&1&0\cr
0&-e^{+i(\alpha-\beta)}&0&0\cr}\pmatrix{c_{00} (-p)\cr c_{11} (-p)\cr
c_{10} (-p)\cr c_{1-1} (-p)\cr}\,.
\end{eqnarray}
This is compatible with the Majorana-like anzatzen.
Of course, the same procedure can be applied in the construction of the quantum field operator for $F_{\mu\nu}$.

%\subsection{The $(1,0)\oplus (0,1)$ Representation.}

The solutions of the Weinberg-like equation
\begin{equation}
[\gamma^{\mu\nu} \partial_\mu \partial_\nu - {(i\partial/\partial t)\over E} 
m^2 ] \Psi (x) =0\,.
\end{equation}
are found in Refs.~\cite{Novozh,Sankar,DVA,DVO94}. Here they are:
\begin{eqnarray}
&&u_\sigma ({\bf p})= \pmatrix{D^S (\Lambda_R) \xi_\sigma ({\bf 0})\cr 
D^S (\Lambda_L)\xi_\sigma ({\bf 0})\cr}\,,\,\,
v_\sigma ({\bf p})=\pmatrix{D^S (\Lambda_R \Theta_{[1/2]}) \xi_\sigma^\ast ({\bf 0})\cr 
- D^S (\Lambda_L \Theta_{[1/2]})\xi_\sigma^\ast ({\bf 0})\cr}=\Gamma^5 u_\sigma ({\bf p}),\nonumber\\ 
&&\\
&&\Gamma^5 = \pmatrix{1_{3\times 3}&0_{3\times 3}\cr
0_{3\times 3}&-1_{3\times 3}\cr}\,,
\end{eqnarray}
where $D^S$ is the matrix of the $(S,0)$ representation of the spinor group $SL (2,c)$. 
In the $(1,0)\oplus (0,1)$ representation 
the procedure of derivation of the creation operators leads to somewhat different situation:
\begin{equation}
\sum_{\sigma=0,\pm 1}^{} v_\sigma (p) b_\sigma^\dagger (p) = \sum_{\sigma=0,\pm 1}^{} u_\sigma (-p) a_\sigma (-p)\,,\quad 
\mbox{hence}\,\,b_\sigma^\dagger (p) = 0\,.
\end{equation}
However, if we return to the original Weinberg equations $[\gamma^{\mu\nu} \partial_\mu \partial_\nu \pm 
m^2 ] \Psi_{1,2} (x) =0$ with the field operators:
\begin{eqnarray}
\Psi_1 (x) &=& \frac{1}{(2\pi)^3}\sum_\mu \int \frac{d^3 {\bf p}}{2E_p} [ u_\mu ({\bf p}) a_\mu ({\bf p}) e^{-ip_\mu\cdot x^\mu}
+ u_\mu ({\bf p}) b_\mu^\dagger ({\bf p}) e^{+ip_\mu\cdot x^\mu}],\nonumber\\
&&\\
\Psi_2 (x) &=& \frac{1}{(2\pi)^3}\sum_\mu \int \frac{d^3 {\bf p}}{2E_p} [ v_\mu ({\bf p}) c_\mu ({\bf p}) e^{-ip_\mu\cdot x^\mu}
+ v_\mu ({\bf p}) d^\dagger_\mu ({\bf p}) e^{+ip_\mu\cdot x^\mu}],
\end{eqnarray}
we obtain
\begin{eqnarray}
b_\mu^\dagger (p) &=& [1-2({\bf S}\cdot {\bf n})^2]_{\mu\lambda} a_\lambda (-p)\,,\label{ba}\\
d_\mu^\dagger (p) &=& [1-2({\bf S}\cdot {\bf n})^2]_{\mu\lambda} c_\lambda (-p)\,\label{ca}.
\end{eqnarray}
The applications of $\overline u_\mu (-p) u_\lambda (-p) = \delta_{\mu\lambda}$ and 
$\overline u_\mu (-p) u_\lambda (p) = [1- 2({\bf S}\cdot {\bf n})^2]_{\mu\lambda}$ prove that the equations
are self-consistent.
This situation signifies that in order to construct the Sankaranarayanan-Good field operator (which was used by Ahluwalia, Johnson and 
Goldman~\cite{DVA}) we need additional postulates. One can try to construct 
the left- and the right-hand side of the field operator separately each other.
In this case the commutation relations may also be more complicated.

Repeating the above procedures, 
on using (\ref{ba}) and the Majorana postulate, we come to:
\begin{equation}
a_\mu^\dagger (p) = + e^{+i\varphi} [1-2({\bf S}\cdot {\bf n})^2]_{\mu\lambda} a_\lambda (-p)\,.\label{ma1S1}
\end{equation}
On the other hand, on using the inverse relation, namely, that for $a_\mu (-p)$, we make the substitutions $E_p \rightarrow -E_p$, ${\bf p} 
\rightarrow -{\bf p}$ to obtain 
\begin{equation}
a_\mu (p) = + [1-2({\bf S}\cdot {\bf n})^2]_{\mu\lambda} b_\lambda^\dagger (-p)\,.\label{ma2S1}
\end{equation}
The totally reflected Majorana anzatz is $b_\mu (-E_p, -{\bf p}) = e^{i\varphi} a_\mu (-E_p, -{\bf p})$. Thus,
\begin{equation}
b_\mu^\dagger (- p) = e^{-i\varphi} a_\mu^\dagger (- p)\,.
\end{equation}
Combining with (\ref{ma2S1}),  we come to
\begin{equation}
a_\mu (p) = + e^{-i\varphi} [1-2({\bf S}\cdot {\bf n})^2]_{\mu\lambda} a_\lambda^\dagger (-p)\,,
\end{equation}
and 
\begin{equation}
a_\mu^\dagger (p) = +e^{+i\varphi} [1-2({\bf S}^\ast \cdot {\bf n})^2]_{\mu\lambda} a_\lambda (-p)\,.
\end{equation}
In the basis where $S_z$ is diagonal the matrix $S_y$ is imaginary~\cite{Var}. So, $({\bf S}^\ast \cdot {\bf n})= S_x n_x - S_y n_y + S_z n_z$, and $({\bf S}^\ast\cdot {\bf n})^2 \neq ({\bf S}\cdot {\bf n})^2$ in the case of $S=1$. So, we conclude that there is the same problem in this 
point, in the aplication of the Majorana-like anzatz, as in the case of spin-1/2. Similarly, one can proceed with (\ref{ca}).

Meanwhile, the attempts of constructing the self/anti-self charge conjugate states failed in Ref.~\cite{DVA1995}. Instead, the $\Gamma^5 S^c_{[1]}-$ self/anti-self conjugate states
have been constructed therein.

%\newpage

\section{Conclusions.}

We conclude that something is missed in the foundations of both the Weyl theory, the original Majorana theory and its generalizations.
Similar problems exist in the theories of higher spins.

\medskip

{\bf Acknowledgements.} I acknowledge discussions with colleagues at recent conferences.
%I am grateful to the Zacatecas University for professorship. 
%The author declares that there is no conflict of interest regarding the publication of this paper.
}
\smallskip

%\bigskip

%\section*{References}

\end{document}